\documentclass[english,notitlepage,pra,twocolumn]{revtex4-1}
\usepackage[T1]{fontenc}
\usepackage[latin9]{inputenc}
\usepackage{amsmath}
\usepackage{amssymb}
\usepackage{graphicx}
\usepackage{babel}
\begin{document}
\title{Universal scattering phase shift in the presence of spin-orbit coupling}
\author{Cai-Xia Zhang$^{1,3}$and Shi-Guo Peng$^{2}$}
\email{pengshiguo@wipm.ac.cn}

\affiliation{$^{1}$Guangdong Provincial Key Laboratory of Quantum Engineering
and Quantum Materials, GPETR Center for Quantum Precision Measurement
and SPTE, South China Normal University, Guangzhou 510006, China}
\affiliation{$^{2}$State Key Laboratory of Magnetic Resonance and Atomic and Molecular
Physics, Innovation Academy for Precision Measurement Science and
Technology, Chinese Academy of Sciences, Wuhan 430071, China}
\affiliation{$^{3}$Guangdong-Hong Kong Joint Laboratory of Quantum Matter, Frontier
Research Institute for Physics, South China Normal University, Guangzhou
510006, China}
\date{\today}
\begin{abstract}
Scattering phase shift, as a key parameter in scattering theory, plays
an important role in characterizing low-energy collisions between
ultracold atoms. In this work, we theoretically investigate the universal
low-energy behavior of the scattering phase shifts for cold atoms
in the presence of spin-orbit coupling. We first construct the asymptotic
form of the two-body wave function when two fermions get as close
as the interaction range, and consider perturbatively the correction
of the spin-orbit coupling up to the second order, in which new scattering
parameters are introduced. Then for elastic collisions, the scattering
phase shifts are defined according to the unitary scattering $S$
matrix. We show how the low-energy behavior of the scattering phase
shifts is modified by these new scattering parameters introduced by
spin-orbit coupling. The universality of the scattering phase shifts
is manifested as the independence of the specific form of the interatomic
potential. The explicit forms of the new scattering parameters are
analytically derived within a model of the spherical-square-well potential.
Our method provides a unified description of the low-energy properties
of scattering phase shifts in the presence of spin-orbit coupling.
\end{abstract}
\maketitle

\section{Introduction}

Owing to their versatility, ultracold atomic gases provide ideal platform
on which to study fascinating quantum many-body phenomena in a highly
controllable and tunable way \cite{Bloch2008M,Giorgini2008T,Chin2010F}.
As a building block of interacting many-body systems, the two-body
problem is of fundamental importance in ultracold atomic physics \cite{Braaten2006U,Blume2012F,Greene2017U}.
In one hand, two-body solutions determine the essential interaction
parameter in the many-body Hamiltonian. In the other hand, the two-body
physics even gives rise to a set of universal relations that characterize
various properties of many-body systems, ranging from macroscopic
thermodynamics to microscopic correlation functions \cite{Tan2008E,Tan2008L,Tan2008G,Zwerger2011T}.
It then opens up a new direction of studying many-body problems based
on the two-body physics \cite{Braaten2008E,Zhang2009U,Stewart2010V,Kuhnle2010U,Yu2015U,Yoshida2015U,He2016C,Luciuk2016E,Peng2016L,Peng2019U}.
An important feature of ultracold atomic systems is that the mean
distance between atoms is usually much larger than the length scale
associated with interatomic potentials. Therefore, the two-body scattering
properties outside the interatomic potential become independent of
the short-range detail of the potential between atoms, and are universally
characterized by the so-called scattering phase shift \cite{Landau1999Q}.
Moreover, microscopic two-body scattering parameters, such as the
scattering length and effective range, can be defined based on the
low-energy expansion of the scattering phase shift \cite{Mott1965T}.

Recently, the successful attainment of the spin-orbit (SO) coupling
in cold atomic gases is one of remarkable breakthroughs \cite{Lin2011S,Cheuk2012S,Wang2012S,Wu2016R,Huang2016E,Chen2018S,Chen2018R,Zhang2019G}.
Owing to the high controllability of cold atoms in interatomic interaction,
purity, and geometry \cite{Chin2010F,Kohler2006P,Zhu2006S,Peng2014M,Gross2017Q},
SO-coupled cold atoms have became a manifold platform for further
researching and understanding of novel phenomena in condensed-matter
physics, such as topological insulators and superconductors \cite{Qi2011T,Dalibard2011C,Zhang12019T}.
However, the introducing of SO coupling brings new challenges to the
few-body physics as well as many-body physics. In the presence of
SO coupling, the center-of-mass (c.m.) motions of pairs are coupled
to their relative motions \cite{Wang2016T}. Besides, all the scattering
partial waves are mixed, since the orbital angular momentum of the
relative motion of two atoms is no longer conserved because of SO
coupling \cite{Cui2012M}. These complications dramatically affect
the theoretical description of interactions between cold atoms, and
especially modify the asymptotic behavior of the many-body wave function
when two atoms get close \cite{Zhang2012M}. Then new scattering parameters
need to be introduced by SO coupling besides the well-known scattering
length and effective range \cite{Peng2018C,Zhang2020U}. Consequently,
Tan's universal relations, governed by the short-range behavior of
the two-body physics, are also amended by SO-coupling. New contacts
are then introduced, and additional universal relations related to
SO-coupling appear \cite{Peng2018C,Zhang2020U,Zhang2018U,Jie2018U,Qin2020L,Qin2020U}.

Two-body scattering problems in the presence of SO coupling have comprehensively
been studied in the past few years \cite{Cui2012M,Duan2013U,Guan2016S,Guan2016A,Wang2018U,Wu2013S}.
However, a unified description of the low-energy behavior of the scattering
phase shift is still elusive. Since new scattering parameters need
to be introduced by SO coupling \cite{Peng2018C,Zhang2020U}, it is
of special interest how these new scattering parameters characterize
the low-energy behavior of the scattering phase shift, and whether
it is universal and independent of the specific form of interatomic
potentials. It is the target of the present work to address these
issues. In this work, we consider the two-body scattering problem
of Fermi gases in the presence of three-dimensional (3D) isotropic
SO coupling. We perturbatively take into account of SO coupling up
to the second-order correction, and construct the short-range asymptotic
form of the two-body wave function. We show that additional scattering
parameters need to be introduced by SO coupling, besides the scattering
length (volume) and effective range. Utilizing the unitarity of the
elastic collision between two fermions, we define the scattering phase
shifts according to the $S$ matrix, and then represent the form of
the two-body wave function outside the interatomic potential by the
scattering phase shifts. By expanding the two-body wave function at
short distance, and comparing it with that obtained perturbatively
by introducing new scattering parameters, we acquire the low-energy
behavior of the scattering phase shifts. The low-energy expansion
of the scattering phase shifts, characterizing by new scattering parameters,
is found universal and independent of the specific form of interatomic
potentials. As a simple example, we verify our results within a model
of the spherical-square-well potential. The explicit forms of new
scattering parameters introduced by SO coupling are analytically derived,
which show excellent agreement with previous numerical calculations
near the $s$-wave resonance \cite{Cui2012M}.

The paper is arranged as follows. To warm up, we briefly review our
method in Sec.II to construct the short-range asymptotic form of the
two-body wave function in the presence of SO coupling, and take into
account of the second-order corrections of SO coupling. In Sec.III,
the scattering phase shifts are defined according to the unitary $S$
matrix for elastic collisions, based on the exact two-body solution
outside the interatomic potential. Then the asymptotic form of the
two-body wave function represented by the scattering phase shifts
is obtained. By comparing this short-range form of the two-body wave
function with that obtained perturbatively by introducing new scattering
parameters, the low-energy behavior of the scattering phase shifts
is then acquired in Sec.IV. The explicit forms of the new scattering
parameters introduced by SO coupling are derived and verified by using
the model of a spherical-square-well potential in Sec.V. Finally,
the remarks and conclusions are summarized in Sec.VI.

\section{model and two-body wave function }

To generalize our previous results \cite{Peng2018C,Zhang2020U}, we
briefly review the introducing of new parameters in handling two-body
scattering problems with SO coupling, and construct the short-range
form of the two-body wave function up to the second-order corrections
of SO coupling. Let us consider two spin-$1/2$ fermions in the presence
of 3D isotropic SO coupling, and the single-particle Hamiltonian takes
the form of \cite{Cui2012M,Peng2018C,Wu2013S}

\begin{equation}
\hat{\mathcal{H}}_{1}=\frac{\hbar^{2}\mathbf{\hat{k}^{\mathrm{2}}}}{2M}+\frac{\hbar^{2}\lambda}{M}\mathbf{\hat{k}}\cdot\hat{\boldsymbol{\sigma}}+\frac{\hbar^{2}\lambda^{2}}{2M},\label{eq:t1}
\end{equation}
where $\hat{{\bf k}}=-i\nabla$ and $\hat{\boldsymbol{\sigma}}$ are
respectively the single-particle momentum and spin operators, $\lambda>0$
denotes the strength of SO coupling, $M$ is the atomic mass, and
$\hbar$ is the Planck's constant divided by $2\pi$. For two fermions,
by introducing the c.m. and relative coordinates ${\bf R}=\left({\bf r}_{1}+{\bf r}_{2}\right)/2$
and ${\bf r}={\bf r}_{1}-{\bf r}_{2}$ as usual, the two-body Hamiltonian
can formally be written as $\hat{\mathcal{H}}_{2}=\hat{\mathcal{H}}_{cm}+\hat{\mathcal{H}}_{r}$
with

\begin{eqnarray}
\mathcal{\hat{H}}_{cm} & = & \frac{\hbar^{2}\hat{\mathbf{K}}^{2}}{4M}+\frac{\hbar^{2}\lambda}{2M}\mathbf{\hat{K}}\cdot\left(\hat{\boldsymbol{\sigma}}_{1}+\hat{\boldsymbol{\sigma}}_{2}\right),\label{eq:t3}\\
\mathcal{\hat{H}}_{r} & = & \frac{\hbar^{2}\mathbf{\hat{k}}^{2}}{M}+\frac{\hbar^{2}\lambda}{M}\mathbf{\hat{k}}\cdot\left(\hat{\boldsymbol{\sigma}}_{1}-\hat{\boldsymbol{\sigma}}_{2}\right)+\frac{\hbar^{2}\lambda^{2}}{M}+V(r),\label{eq:t4}
\end{eqnarray}
which describe the c.m. motion with total momentum $\hat{{\bf K}}$
and relative motion with momentum $\hat{{\bf k}}=\left(\hat{{\bf k}}_{1}-\hat{{\bf k}}_{2}\right)/2$,
respectively. Here, $V\left(r\right)$ is the short-range interaction
potential between two fermions. As discussed in \cite{Cui2012M},
the total angular momentum ${\bf J}$ of two fermions as well as their
total momentum ${\bf K}$ is conserved. We may conveniently focus
on the scattering problem in the subspace of $\mathbf{K=0}$ and $\mathbf{J}=0$.
In this case, the two-body Hamiltonian $\hat{\mathcal{H}}_{2}$ is
simply reduced to $\hat{\mathcal{H}}_{r}$, and only $s$- and $p$-wave
scatterings are involved. Moreover, the subspace of $\mathbf{K=0}$
and $\mathbf{J}=0$ is spanned by two angular orthogonal basis $\left\{ \Omega_{0}\left(\mathbf{\hat{r}}\right),\Omega_{1}\left(\mathbf{\hat{r}}\right)\right\} $
\cite{Cui2012M,Peng2018C,Zhang2020U}

\begin{eqnarray}
\Omega_{0}\left(\mathbf{\hat{r}}\right) & = & Y_{00}\left(\mathbf{\hat{r}}\right)\left|S\right\rangle ,\\
\Omega_{1}\left(\mathbf{\hat{r}}\right) & = & -\frac{i}{\sqrt{3}}\left[Y_{1-1}\left(\mathbf{\hat{r}}\right)\left|\uparrow\uparrow\right\rangle \right.\\
 &  & \left.+Y_{11}\left(\mathbf{\hat{r}}\right)\left|\downarrow\downarrow\right\rangle -Y_{10}\left(\mathbf{\hat{r}}\right)\left|T\right\rangle \right],
\end{eqnarray}
where $Y_{lm}\left(\hat{{\bf r}}\right)$ with angular variable $\hat{{\bf r}}=\left(\theta,\varphi\right)$
for the relative motion of two fermions is the spherical harmonics,
and $\left|S\right\rangle =\left(\left|\uparrow\downarrow\right\rangle -\left|\downarrow\uparrow\right\rangle \right)/\sqrt{2}$
and $\left\{ \left|\uparrow\uparrow\right\rangle ,\left|\downarrow\downarrow\right\rangle ,\left|T\right\rangle =\left(\left|\uparrow\downarrow\right\rangle +\left|\downarrow\uparrow\right\rangle \right)/\sqrt{2}\right\} $
are the singlet and triplet spin states for two fermions, respectively.
Then the two-body wave function in this subspace can generally be
written in the form of

\begin{equation}
\Psi\left(\mathbf{r}\right)=\psi_{0}\left(r\right)\Omega_{0}\left(\mathbf{\hat{r}}\right)+\psi_{1}\left(r\right)\Omega_{1}\left(\mathbf{\hat{r}}\right),\label{eq:t8}
\end{equation}
and $\psi_{i}\left(r\right)\,\left(i=0,1\right)$ denotes the radial
part of the wave function.

The existence of SO coupling dramatically changes the short-range
behavior of the two-body wave function \cite{Zhang2012M}. However,
for a realistic interaction potential $V\left(r\right)$ with a short
range $\epsilon$, the SO-coupling strength $\lambda$ as well as
the relative momentum $k$ between two fermions is usually much smaller
than $\epsilon^{-1}$ in current experiments of cold atoms \cite{Cheuk2012S,Wang2012S}.
In this case, we may perturbatively construct the asymptotic form
of the two-body wave function when two fermions approach as close
as $\epsilon$, i.e.,

\begin{multline}
\Psi\left(\mathbf{r}\right)\approx\phi\left(\mathbf{r}\right)+k^{2}F(\mathbf{r})-\lambda G(\mathbf{r})\\
+k^{4}X(\mathbf{r})-\lambda k^{2}Y\left(\mathbf{r}\right)-\lambda^{2}Z\left(\mathbf{r}\right).\label{eq:aw}
\end{multline}
To generalize our previous results \cite{Peng2018C,Zhang2020U}, here
we take into account of the second-order corrections of $\left(k^{2},\lambda\right)$.
Later we will see that these second-order corrections modify the low-energy
expansion of the scattering phase shifts. 

Substituting Eq.(\ref{eq:aw}) into the two-body Schr\"{o}dinger
equation $\mathcal{\hat{H}}_{2}\Psi\left(\mathbf{r}\right)=E\Psi\left(\mathbf{r}\right)$,
following the similar route as that in \cite{Peng2018C,Zhang2020U},
and after some straightforward algebra, we obtain the general form
of the two-body wave function at short distance (see appendix for
detail) \begin{widetext}
\begin{eqnarray}
\Psi\left(\mathbf{r}\right) & = & \alpha_{0}\left[\frac{1}{r}+\left(-\frac{1}{a_{0}}+\frac{b_{0}k^{2}}{2}+\frac{\alpha_{1}}{\alpha_{0}}u\lambda-\frac{\alpha_{1}}{\alpha_{0}}h\lambda k^{2}-c_{0}\lambda^{2}\right)-\left(\frac{k^{2}}{2}+\frac{\alpha_{1}}{\alpha_{0}}\lambda k^{2}+\frac{3\lambda^{2}}{2}\right)r\right]\Omega_{0}\left(\mathbf{\hat{r}}\right)\nonumber \\
 &  & +\alpha_{1}\left[\frac{1}{r^{2}}+\frac{1}{2}\left(k^{2}+2\lambda\frac{\alpha_{0}}{\alpha_{1}}-\lambda^{2}\right)+\left(-\frac{1}{3a_{1}}+\frac{b_{1}k^{2}}{6}+\frac{\alpha_{0}}{\alpha_{1}}v\lambda-\frac{\alpha_{0}}{\alpha_{1}}q\lambda k^{2}-c_{1}\lambda^{2}\right)r\right]\Omega_{1}\left(\mathbf{\hat{r}}\right)+\mathcal{O}\left(r^{2}\right)\label{eq:ShortRangeWF}
\end{eqnarray}
\end{widetext} for $r\gtrsim\epsilon$, where $a_{0},b_{0}$ ($a_{1},b_{1}$)
are the well-known $s$-wave ($p$-wave) scattering length (volume)
and effective range, $u,v$ are the new scattering parameters resulted
from the first-order correction of SO coupling, and $h,q,c_{0},c_{1}$
are those introduced by the second-order correction of SO coupling.
Here, $\alpha_{0}$ and $\alpha_{1}$ are two complex superposition
coefficients. Near the $s$-wave resonance, the contribution of the
$p$-wave scattering could be ignored and we have $\alpha_{1}\approx0$.
The short-range form of the two-body wave function reduces to (up
to a constant $\alpha_{0}$)
\begin{multline}
\Phi_{0}\left(\mathbf{r}\right)=\left[\frac{1}{r}+\left(-\frac{1}{a_{0}}+\frac{b_{0}k^{2}}{2}-c_{0}\lambda^{2}\right)\right.\\
\left.-\frac{1}{2}\left(k^{2}+3\lambda^{2}\right)r\right]\Omega_{0}\left(\hat{\mathbf{r}}\right)\\
+\lambda\left[1+\left(v-qk^{2}\right)r\right]\Omega_{1}\left(\hat{\mathbf{r}}\right)+\mathcal{O}\left(r^{2}\right),\label{eq:sr}
\end{multline}
which recovers the modified Bethe-Peierls boundary condition of \cite{Zhang2012M},
by noticing that we expand the wave function up to the order of $r$
at short distance and the second-order terms of $\lambda$ are retained.
We can see that a considerable $p$-wave component is involved because
of SO coupling, even near the $s$-wave resonance. In like manner,
near the $p$-wave resonance, the contribution of the $s$-wave scattering
is small and could be ignored ($\alpha_{0}\approx0$). Subsequently,
the form of the two-body wave function at short distance reduces to
(up to a constant $\alpha_{1}$)
\begin{multline}
\Phi_{1}\left(\mathbf{r}\right)=\lambda\left[\left(u-hk^{2}\right)-k^{2}r\right]\Omega_{0}\left(\hat{\mathbf{r}}\right)+\left[\frac{1}{r^{2}}+\frac{1}{2}\left(k^{2}-\lambda^{2}\right)\right.\\
\left.+\left(-\frac{1}{3a_{1}}+\frac{b_{1}k^{2}}{6}-c_{1}\lambda^{2}\right)r\right]\Omega_{1}\left(\hat{\mathbf{r}}\right)+\mathcal{O}\left(r^{2}\right).\label{eq:pr}
\end{multline}
Similarly, an $s$-wave component is induced by SO coupling even near
the $p$-wave resonance.

\section{The $S$ matrix and scattering phase shifts}

The range $\epsilon$ of interaction potentials between neutral atoms
is usually much smaller than the inverse of the relative momentum
$k$ as well as that of SO-coupling strength $\lambda$. The two-body
wave function outside the potential in the presence of SO coupling
can easily be obtained by solving the Schr\"{o}dinger equation with
$V\left(r\right)=0$ for $r>\epsilon$ \cite{Wu2013S}, 
\begin{multline}
\Psi\left({\bf r}\right)=A\Psi_{-}^{(in)}\left({\bf r}\right)+B\Psi_{+}^{(in)}\left({\bf r}\right)\\
+C\Psi_{-}^{(out)}\left({\bf r}\right)+D\Psi_{+}^{(out)}\left({\bf r}\right)\label{eq:sw}
\end{multline}
with
\begin{equation}
\Psi_{\pm}^{(in)}\left({\bf r}\right)=k_{\pm}\left[h_{0}^{(2)}\left(k_{\pm}r\right)\Omega_{0}\left(\hat{\mathbf{r}}\right)\pm h_{1}^{(2)}\left(k_{\pm}r\right)\Omega_{1}\left(\hat{\mathbf{r}}\right)\right],\label{eq:sw1}
\end{equation}
\begin{equation}
\Psi_{\pm}^{(out)}\left({\bf r}\right)=k_{\pm}\left[h_{0}^{(1)}\left(k_{\pm}r\right)\Omega_{0}\left(\hat{\mathbf{r}}\right)\pm h_{1}^{(1)}\left(k_{\pm}r\right)\Omega_{1}\left(\hat{\mathbf{r}}\right)\right],\label{eq:sw2}
\end{equation}
 $k_{\pm}=k\pm\lambda$, and $k=\sqrt{ME/\hbar^{2}}$. Here, $h_{\nu}^{\left(1\right)}$
and $h_{\nu}^{\left(2\right)}$ denote the $\nu$th-order spherical
Hankel functions of the first and second kinds, respectively, and
$A,B,C,D$ are superposition coefficients. The physical meaning of
the solution (\ref{eq:sw}) is apparent: due to unique properties
of the single-particle dispersion relation \cite{Cui2012M,Wu2013S},
the incident wave with energy $E$ is an arbitrary superposition of
two spherical waves with two different magnitudes of momenta $k_{\pm}$
(corresponding to the spherical Hankel function of the second kind
$h_{\nu}^{(2)}$); scattered elastically by the interatomic potential
$V\left(r\right)$, the outgoing wave becomes a different superposition
of the same two spherical waves (corresponding to the spherical Hankel
function of the first kind $h_{\nu}^{(1)}$).

Using the time-dependent Schr\"{o}dinger equation $i\hbar\partial\Psi\left({\bf r},t\right)/\partial t=\hat{\mathcal{H}}_{2}\Psi\left({\bf r},t\right)$,
and noticing $\Psi\left({\bf r},t\right)=\psi_{0}\left(r,t\right)\Omega_{0}\left(\mathbf{\hat{r}}\right)+\psi_{1}\left(r,t\right)\Omega_{1}\left(\mathbf{\hat{r}}\right)$,
we easily obtain the following continuity equation

\begin{equation}
\frac{\partial\rho}{\partial t}+\frac{dj}{dr}=0,\label{eq:s2}
\end{equation}
where $\rho=r^{2}\left(\left|\psi_{0}\right|^{2}+\left|\psi_{1}\right|^{2}\right)$
is the radial probability density, and 
\begin{multline}
j\left(r\right)=\\
\frac{i\hbar}{M}r^{2}\left[\sum_{i=0}^{1}\left(\psi_{i}\frac{d\psi_{i}^{*}}{dr}-\psi_{i}^{*}\frac{d\psi_{i}}{dr}\right)-2\lambda\left(\psi_{0}^{*}\psi_{1}-\psi_{0}\psi_{1}^{*}\right)\right]\label{eq:s3}
\end{multline}
is the radial probability current density. For the stationary state
of the system, the radial probability density $\rho$ is obviously
independent of time, and thus the radial probability current density
$j\left(r\right)$ is a constant. Moreover, we have $j\left(r\right)=0$
at $r=0$, which in turn indicates that $j\left(r\right)$ should
be zero everywhere. Inserting the scattering solution Eq.(\ref{eq:sw})
into Eq.(\ref{eq:s3}) and using $j\left(r\right)=0$, we easily obtain

\begin{equation}
\left|A\right|^{2}+\left|B\right|^{2}=\left|C\right|^{2}+\left|D\right|^{2}.\label{eq:s5}
\end{equation}
This means that the outgoing wave is different from the incident wave
only up to a unitary transformation, i.e.

\begin{equation}
\left[\begin{array}{c}
C\\
D
\end{array}\right]=\mathcal{S}\left[\begin{array}{c}
A\\
B
\end{array}\right],\label{eq:s6}
\end{equation}
and $\mathcal{S}$ is a unitary $2\times2$ matrix determined by the
specific form of the short-range interaction potential. The unitarity
of the matrix $\mathcal{S}$ was confirmed by using a spherical-square-well
model in \cite{Wu2013S}. Here, we emphasize that this property is
obviously universal for any short-range interaction potential in the
presence of SO coupling, which is a natural consequence of the probability
conservation during elastic collisions.

The unitary $\mathcal{S}$ matrix can formally be diagonalized as
\begin{equation}
W^{\dagger}\mathcal{S}W=\left[\begin{array}{cc}
e^{i2\delta_{0}} & 0\\
0 & e^{i2\delta_{1}}
\end{array}\right]
\end{equation}
by
\begin{equation}
W=\left[\begin{array}{cc}
\omega_{0-} & \omega_{1-}\\
\omega_{0+} & \omega_{1+}
\end{array}\right],
\end{equation}
where $\left[\omega_{0-},\omega_{0+}\right]^{T}$ and $\left[\omega_{1-},\omega_{1+}\right]^{T}$
are two eigenvectors of $\mathcal{S}$ corresponding to different
eigenvalues $e^{i2\delta_{0}}$ and $e^{i2\delta_{1}}$. We can see
that $\delta_{0,1}$ are the new scattering phase shifts characterizing
the scattering effect in the presence of SO coupling. Inserting Eq.(\ref{eq:s6})
into Eq.(\ref{eq:sw}), we may represent the solution $\Psi\left({\bf r}\right)$
outside the interatomic potential by using eigenvectors and eigenvalues
of the $\mathcal{S}$ matrix, i.e., 
\begin{equation}
\Psi\left({\bf r}\right)=\alpha_{0}\Phi_{0}\left({\bf r}\right)+\alpha_{1}\Phi_{1}\left({\bf r}\right)\label{eq:sp2}
\end{equation}
 with
\begin{multline}
\Phi_{i}\left({\bf r}\right)=\omega_{i-}\Psi_{-}^{(in)}+\omega_{i+}\Psi_{+}^{(in)}\\
+e^{i2\delta_{i}}\left(\omega_{i-}\Psi_{-}^{(out)}+\omega_{i+}\Psi_{+}^{(out)}\right),\label{eq:ns}
\end{multline}
and $\alpha_{i}=\omega_{i-}^{*}A+\omega_{i+}^{*}B$ for $i=0,1$.
Substituting Eqs. (\ref{eq:sw1}) and (\ref{eq:sw2}) into Eq.(\ref{eq:ns}),
we arrive at

\begin{multline}
\Phi_{i}\left({\bf r}\right)=2e^{i\delta_{i}}\cos\delta_{i}\\
\times\sum_{\nu=0}^{1}\sum_{\eta=\pm}\eta^{\nu}\omega_{i\eta}k_{\eta}\left[j_{\nu}\left(k_{\eta}r\right)-n_{\nu}\left(k_{\eta}r\right)\tan\delta_{i}\right]\Omega_{\nu}\left(\hat{{\bf r}}\right)\label{eq:ss}
\end{multline}
for $i=0,1$, where $j_{\nu}$ and $n_{\nu}$ are the $\nu$th-order
spherical Bessel functions of the first and second kinds, respectively. 

\section{low energy expansion of scattering phase shifts}

For ultracold atoms, the low-energy behavior of the scattering phase
shift gives rise to some key microscopic scattering parameters, such
as the scattering length (volume) and effective range. In the presence
of SO coupling, the low-energy expansion of the scattering phase shift
is expected to be modified. Then we may consider such modification
perturbatively, since the energy scale of the SO-coupling strength
is usually much smaller than that corresponding to the range of interatomic
potentials. This can be done by simply comparing the short-range expansion
of Eq.(\ref{eq:ss}) with those of Eqs.(\ref{eq:sr}) and (\ref{eq:pr}).

Near $s$-wave resonances, we may expand Eq.(\ref{eq:ss}) with $i=0$
at small $r$ and obtain

\begin{multline}
\Phi_{0}\left(\mathbf{r}\right)=\left(\frac{1}{r}+\frac{k+\omega_{0}\lambda}{\tan\delta_{0}}-\frac{k^{2}+\lambda^{2}+2\omega_{0}\lambda k}{2}r\right)\Omega_{0}\left(\hat{\mathbf{r}}\right)\\
+\left[\frac{\omega_{0}k-\lambda}{k^{2}-\lambda^{2}}\frac{1}{r^{2}}+\frac{\omega_{0}k+\lambda}{2}\right.\\
\left.+\frac{\omega_{0}\left(k^{2}+\lambda^{2}\right)+2\lambda k}{3\tan\delta_{0}}r\right]\Omega_{1}\left(\hat{\mathbf{r}}\right)+\mathcal{O}\left(r^{2}\right)\label{eq:srr}
\end{multline}
for $r\sim\epsilon^{+}$, where we have omitted an overall factor
$2\left(\omega_{0+}+\omega_{0-}\right)e^{i\delta_{0}}\sin\delta_{0}$,
and introduced $\omega_{0}\equiv\left(\omega_{0+}-\omega_{0-}\right)/\left(\omega_{0+}+\omega_{0-}\right)$.
Comparing the corresponding terms between the Eq.(\ref{eq:srr}) and
Eq.(\ref{eq:sr}), we find
\begin{equation}
k\cot\delta_{0}\approx\frac{k^{2}}{k^{2}+\lambda^{2}}\left(-\frac{1}{a_{0}}+\frac{b_{0}}{2}k^{2}-c_{0}\lambda^{2}\right),\label{eq:s0}
\end{equation}
which recovers that of \cite{Cui2012M} if we keep the terms in $\tan\delta_{0}$
up to $\lambda^{2}$ and notice $b_{0}\approx0$ near broad $s$-wave
resonances. It is apparent that $\delta_{0}$ is the counterpart of
the $s$-wave scattering phase shift in the absence of SO coupling,
and then Eq.(\ref{eq:s0}) reduces to the well-known effective-range
expansion of the $s$-wave scattering phase shift, i.e., $k\cot\delta_{0}=-1/a_{0}+b_{0}k^{2}/2+\mathcal{O}\left(k^{4}\right)$.
In the presence of SO coupling, the new scattering parameter $c_{0}$
is involved, which modifies the low-energy behavior of the $s$-wave
scattering phase shift.

Near $p$-wave resonances, we again expand Eq.(\ref{eq:ss}) with
$i=1$ at small $r$ and obtain (up to an overall constant $2\left(\omega_{1+}/k_{+}-\omega_{1-}/k_{-}\right)e^{i\delta_{1}}\sin\delta_{1}$)

\begin{multline}
\Phi_{1}\left(\mathbf{r}\right)=\left[\frac{\omega_{1}\left(k^{2}-\lambda^{2}\right)}{k-\omega_{1}\lambda}\frac{1}{r}+\frac{\left(k^{2}-\lambda^{2}\right)\left(\omega_{1}k+\lambda\right)}{\left(k-\omega_{1}\lambda\right)\tan\delta_{1}}\right.\\
\left.-\frac{\omega_{1}\left(k^{4}-\lambda^{4}\right)+2\lambda k^{3}-2\lambda^{3}k}{2\left(k-\omega_{1}\lambda\right)}r\right]\Omega_{0}\left(\hat{\mathbf{r}}\right)\\
+\left[\frac{1}{r^{2}}+\frac{\left(k^{2}-\lambda^{2}\right)\left(k+\omega_{1}\lambda\right)}{2\left(k-\omega_{1}\lambda\right)}+\frac{k^{2}-\lambda^{2}}{3\left(k-\omega_{1}\lambda\right)}\right.\\
\left.\times\frac{k^{2}+\lambda^{2}+2\omega_{1}\lambda k}{\tan\delta_{1}}r\right]\Omega_{1}\left(\hat{\mathbf{r}}\right)+\mathcal{O}(r^{2})\label{eq:prr}
\end{multline}
for $r\sim\epsilon^{+}$, and $\omega_{1}\equiv\left(\omega_{1+}+\omega_{1-}\right)/\left(\omega_{1+}-\omega_{1-}\right)$.
In like manner, comparing the corresponding terms of Eq.(\ref{eq:prr})
with Eq.(\ref{eq:pr}), we easily obtain 

\begin{equation}
k^{3}\cot\delta_{1}\approx\frac{k^{4}}{k^{4}-\lambda^{4}}\left(-\frac{1}{a_{1}}+\frac{b_{1}}{2}k^{2}-3c_{1}\lambda^{2}\right).\label{eq:ps}
\end{equation}
It recovers that of \cite{Cui2012M} near $p$-wave resonances \cite{NOTE-PWave}.
We can see that the SO-coupling induced scattering parameter $c_{1}$
is involved in characterizing the scattering phase shift $\delta_{1}$.
Obviously, the scattering phase shift $\delta_{1}$ is the counterpart
of the $p$-wave scattering phase shift in the absence of SO coupling.
It reduces to the well-known effective-range expansion of the $p$-wave
scattering phase shift without SO coupling, i.e., $k^{3}\cot\delta_{1}=-1/a_{1}+b_{1}k^{2}/2+\mathcal{O}\left(k^{4}\right)$. 

\section{a spherical-square-well potential model}

Till now, we have discussed the introducing of new scattering parameters
in the short-range asymptotic behavior of the two-body wave function,
and considered how these new scattering parameters characterize and
modify the scattering phase shifts at low energy. In order to demonstrate
that these new scattering parameters $c_{0}$ and $c_{1}$ are not
artificially introduced by our perturbation method and are in fact
physically determined by real interatomic potentials, in the follows,
we consider a two-body scattering problem within a model of the spherical-square-well
potential, i.e.,

\begin{equation}
V\left(r\right)=\begin{cases}
-V_{0}, & 0\leq r\leq\epsilon,\\
0, & r>\epsilon
\end{cases}\label{eq:sw2-1}
\end{equation}
with the depth $V_{0}>0$. By solving the Schr\"{o}dinger equation
inside and outside the potential, respectively, and utilizing the
continuity of the two-body wave function at $r=\epsilon$ as well
as its first-order derivative, we obtain (see appendix) \begin{widetext}

\begin{eqnarray}
\frac{c_{0}}{\epsilon} & = & -1+\frac{\widetilde{V}_{0}}{3\left[\sqrt{\widetilde{V}_{0}}-\tan\left(\sqrt{\widetilde{V}_{0}}\right)\right]^{2}}+\frac{1}{\widetilde{V}_{0}-\sqrt{\widetilde{V}_{0}}\tan\left(\sqrt{\widetilde{V}_{0}}\right)},\label{eq:sw43}\\
c_{1}\epsilon & = & -\frac{\widetilde{V}_{0}\left[\left(-15+\widetilde{V}_{0}\right)\sqrt{\widetilde{V}_{0}}\cos\left(\sqrt{\widetilde{V}_{0}}\right)+3\left(5-2\widetilde{V}_{0}\right)\sin\left(\sqrt{\widetilde{V}_{0}}\right)\right]}{5\left[-3+\widetilde{V}_{0}+3\sqrt{\widetilde{V}_{0}}\cot\left(\sqrt{\widetilde{V}_{0}}\right)\right]^{2}\left[\sqrt{\widetilde{V}_{0}}\cos\left(\sqrt{\widetilde{V}_{0}}\right)-\sin\left(\sqrt{\widetilde{V}_{0}}\right)\right]},\label{eq:sw44}
\end{eqnarray}
and $\tilde{V}_{0}=M\epsilon^{2}V_{0}/\hbar^{2}$.\end{widetext}

\begin{figure}
\begin{centering}
\includegraphics[bb=23bp 302bp 381bp 578bp,clip,width=7.5cm,height=6cm]{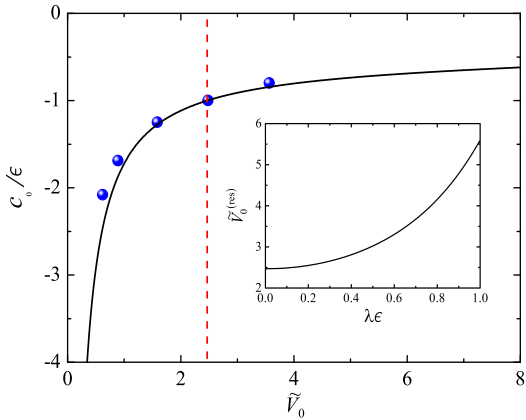}
\par\end{centering}
\caption{(Color online) The new scattering parameter $c_{0}/\epsilon$, characterizing
the second-order correction of $\lambda$ to the scattering phase
shift $\delta_{0}$, evolving with the depth of the potential $\tilde{V}_{0}=M\epsilon^{2}V_{0}/\hbar^{2}$.
The red dashed line indicates the $s$-wave resonance in the absence
of spin-orbit coupling, and the blue numerical data is from \cite{Cui2012M}.
The inset shows the position of $s$-wave scattering resonance changing
with the spin-orbit-coupling strength.}

\label{fig.s-wave}
\end{figure}

\begin{figure}
\begin{centering}
\includegraphics[bb=25bp 290bp 380bp 576bp,clip,width=7.5cm,height=6.5cm]{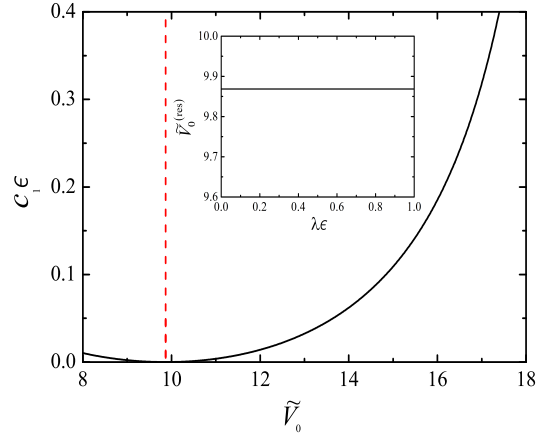}
\par\end{centering}
\caption{(Color online) The new scattering parameter $c_{1}\epsilon$, characterizing
the second-order correction of $\lambda$ to the scattering phase
shift $\delta_{1}$, evolving with the depth potential $\tilde{V}_{0}=M\epsilon^{2}V_{0}/\hbar^{2}$.
The red dashed line indicates the $p$-wave resonance in the absence
of spin-orbit coupling. The inset shows the position of $p$-wave
scattering resonance changing with the spin-orbit-coupling strength. }

\label{fig.p-wave}
\end{figure}

We present $c_{0}$ as a function of the reduced depth $\tilde{V}_{0}$
of the potential near the $s$-wave resonance in Fig.\ref{fig.s-wave}.
The new parameter $c_{0}$ introduced by SO coupling characterizes
the second-order correction of $\lambda$ to the scattering phase
shift $\delta_{0}$, and has numerically been evaluated in \cite{Cui2012M}
near the $s$-wave resonance. Our analytical result, i.e., Eq.(\ref{eq:sw43}),
shows accurate agreement with the numerical calculation of \cite{Cui2012M}.
The scattering phase shift $\delta_{0}$ involves all short-range
information of the interatomic potential, and governs the $s$-wave
scattering properties outside the potential $r>\epsilon$. The resonance
position is determined by $\delta_{0}=\pi/2$, which yields the well
depth $\tilde{V}_{0}=\pi^{2}/4$ at the resonance in the absence of
SO coupling as indicated by the red dashed line in Fig.\ref{fig.s-wave}.
When the SO coupling is gradually turned on, the resonance position
is dramatically shifted by SO coupling because of considerable value
of $c_{0}$, as shown in the inset of Fig.\ref{fig.s-wave}. For the
same consideration, we plot $c_{1}$ as a function of the reduced
depth $\tilde{V}_{0}$ near the $p$-wave resonance in Fig.\ref{fig.p-wave},
which characterizes the second-order correction of $\lambda$ to the
$p$-wave scattering phase shift $\delta_{1}$. Unlike the case of
the $s$-wave scattering, we can see that $c_{1}$ is extremely small
near the $p$-wave resonance without SO coupling, which means the
correction from the SO coupling to the $p$-wave scattering phase
shift is negligibly small. Consequently, the resonance position is
nearly unchanged when the SO coupling is gradually turned on as shown
in the inset of Fig.\ref{fig.p-wave}.

\section{Conclusions}

In this work, we investigate the two-body scattering problems of Fermi
gases in the presence of the three-dimensional isotropic spin-orbit
coupling. Since the energy scale corresponding to the spin-orbit-coupling
strength is much smaller than that corresponding to the range of interatomic
potentials, we perturbatively construct the asymptotic behavior of
two-body wave function when two fermions approach as close as the
interaction range. To generalize our previous results \cite{Peng2018C,Zhang2020U},
we consider up to the second-order correction of the spin-orbit-coupling
strength to the two-body wave function, and additional new scattering
parameters are introduced. Further, the scattering phase shifts are
defined based on the unitary $S$ matrix for an elastic collision.
The low-energy behavior of the scattering phase shifts is discussed,
which is modified by the new scattering parameters in the presence
of spin-orbit coupling. Our results naturally reduce to the well-known
effective-range expansions of the scattering phase shifts without
spin-orbit coupling. Within the model of a spherical-square-well potential,
our analytical results are verified and agree well with previous numerical
calculations near the $s$-wave resonance \cite{Cui2012M}. To simplify
the presentation of this work, we focus our discussions on the subspace
of zero center-of-mass momentum and zero total angular momentum of
two fermions, and then only $s$- and $p$-wave scatterings are involved.
The advantage of our method is that the correction of the spin-orbit
coupling to the two-body wave function at short distance may perturbatively
be considered order by order, as well as to the scattering phase shifts
at low energy. Therefore, it is straightforward for the generalization
to the case of non-zero center-of-mass momentum and non-zero total
angular momentum. Then more scattering partial waves should be involved,
and additional scattering parameters would be introduced to characterize
the short-range behavior of the two-body wave function as well as
the low-energy expansion of the scattering phase shifts. Our method
provides a unified description of the low-energy properties of scattering
phase shifts in the presence of spin-orbit coupling.
\begin{acknowledgments}
C.-X.Z. is supported by NSFC under Grant No.11947097. S.-G.P. is supported
by NSFC under Grants No. 11974384, and NKRDP under Grant No. 2016YFA0301503.
\end{acknowledgments}

\appendix

\section{Scattering under a spherical-square-well potential }

In this appendix, we present the derivation detail of the short-range
behavior of the two-body wave function (\ref{eq:ShortRangeWF}), and
calculate the specific forms of the new scattering parameters for
a spherical-square-well potential. Substituting the ansatz Eq.(\ref{eq:aw})
into the two-body Schr$\ddot{\mathrm{o}}$dinger equation $\mathcal{\hat{H}}_{2}\Psi\left(\mathbf{r}\right)=E\Psi\left(\mathbf{r}\right)$,
and comparing the corresponding coefficients of the terms ( $k^{2}$,
$\lambda$, $k^{4}$, $\lambda k^{2}$, $\lambda^{2}$) on both sides,
we obtain the following coupled equations,

\begin{eqnarray}
\left[-\nabla^{2}+\frac{MV\left(\mathbf{r}\right)}{\hbar^{2}}\right]\phi\left(\mathbf{r}\right) & = & 0,\label{eq:a1}\\
\left[-\nabla^{2}+\frac{MV\left(\mathbf{r}\right)}{\hbar^{2}}\right]F\left(\mathbf{r}\right) & = & \phi\left(\mathbf{r}\right),\label{eq:a2}\\
\left[-\nabla^{2}+\frac{MV\left(\mathbf{r}\right)}{\hbar^{2}}\right]G\left(\mathbf{r}\right) & = & \hat{Q}\left(\mathbf{r}\right)\phi\left(\mathbf{r}\right),\label{eq:a3}\\
\left[-\nabla^{2}+\frac{MV\left(\mathbf{r}\right)}{\hbar^{2}}\right]Y\left(\mathbf{r}\right) & = & \hat{Q}\left(\mathbf{r}\right)F\left(\mathbf{r}\right)+G\left(\mathbf{r}\right),\label{eq:a5}\\
\left[-\nabla^{2}+\frac{MV\left(\mathbf{r}\right)}{\hbar^{2}}\right]Z\left(\mathbf{r}\right) & = & \phi\left(\mathbf{r}\right)-\hat{Q}\left(\mathbf{r}\right)G\left(\mathbf{r}\right)\label{eq:a6}
\end{eqnarray}
with $\hat{Q}\left(\mathbf{r}\right)=\mathbf{\hat{k}}\cdot\left(\hat{\boldsymbol{\sigma}}_{1}-\hat{\boldsymbol{\sigma}}_{2}\right)$.
The corresponding expressions of functions $\phi\left(\mathbf{r}\right)$,
$F(\mathbf{r})$, $G(\mathbf{r})$, $Y\left(\mathbf{r}\right)$ and
$Z\left(\mathbf{r}\right)$ outside the potential for $r\gtrsim\epsilon$
can easily be obtained by solving these coupled equations\begin{widetext}

\begin{eqnarray}
\phi\left(\mathbf{r}\right) & = & \alpha_{0}\left(\frac{1}{r}-\frac{1}{a_{0}}\right)\Omega_{0}\left(\mathbf{\hat{r}}\right)+\alpha_{1}\left(\frac{1}{r^{2}}-\frac{1}{3a_{1}}r\right)\Omega_{1}\left(\mathbf{\hat{r}}\right)+\mathcal{O}\left(r^{2}\right),\label{eq:A1}\\
F\left(\mathbf{r}\right) & = & \alpha_{0}\left(\frac{1}{2}b_{0}-\frac{1}{2}r\right)\Omega_{0}\left(\mathbf{\hat{r}}\right)+\alpha_{1}\left(\frac{1}{2}+\frac{b_{1}}{6}r\right)\Omega_{1}\left(\mathbf{\hat{r}}\right)+\mathcal{O}\left(r^{2}\right),\label{eq:A2}\\
G\left(\mathbf{r}\right) & = & -\alpha_{1}u\Omega_{0}\left(\mathbf{\hat{r}}\right)-\alpha_{0}\left(1+vr\right)\Omega_{1}\left(\mathbf{\hat{r}}\right)+\mathcal{O}\left(r^{2}\right),\label{eq:A3}\\
Y\left(\mathbf{r}\right) & = & \alpha_{1}\left(h+r\right)\Omega_{0}\left(\mathbf{\hat{r}}\right)+\alpha_{0}qr\Omega_{1}\left(\mathbf{\hat{r}}\right)+\mathcal{O}\left(r^{2}\right),\label{eq:A4}\\
Z\left(\mathbf{r}\right) & = & \alpha_{0}\left(c_{0}+\frac{3}{2}r\right)\Omega_{0}\left(\mathbf{\hat{r}}\right)+\alpha_{1}\left(\frac{1}{2}+c_{1}r\right)\Omega_{1}\left(\mathbf{\hat{r}}\right)+\mathcal{O}\left(r^{2}\right).\label{eq:A5}
\end{eqnarray}
\end{widetext}Inserting these functions into the two-body wave function
(\ref{eq:aw}), we arrive at Eq.(\ref{eq:ShortRangeWF}). Obviously,
the derivation here is independent of the specific form of the interatomic
potential, and new scattering parameters are introduced in the short-range
form of the two-body wave function. 

To determine the new scattering parameters in the two-body wave function,
let us consider a specific model of the spherical-square-well potential.
Outside the potential, i.e., $V\left(r\right)=0$, we have already
obtain the form of the two-body wave function as shown in Eqs.(\ref{eq:A1})
to (\ref{eq:A5}). While inside the potential, we have $V\left(r\right)=-V_{0}$,
and the corresponding specific forms of these functions inside the
potential are obtained by solving the Schr\"{o}ding equation. By
using the continuity of the two-body wave function at $r=\epsilon$
as well as its first-order derivative, all the scattering parameters
are then be determined. To simplify the presentation, we only show
the specific forms of $c_{0}$ and $c_{1}$ in Eqs.(\ref{eq:sw43})-(\ref{eq:sw44}),
which characterize the second-order corrections of SO coupling to
the scattering phase shifts.


\begin{thebibliography}{10}
\bibitem{Bloch2008M}I. Bloch, J. Dalibard, and W. Zwerger, Many-body
physics with ultracold gases, Rev. Mod. Phys. $\mathbf{80}$, 885
(2008). 

\bibitem{Giorgini2008T} S. Giorgini, L. P. Pitaevskii, and S. Stringari,
Theory of ultracold atomic Fermi gases, Rev. Mod. Phys. $\mathbf{80}$,
1215 (2008).

\bibitem{Chin2010F}C. Chin, R. Grimm, P. Julienne, and E. Tiesinga,
Feshbach resonances in ultracold gases, Rev. Mod. Phys. $\mathbf{82}$,
1225 (2010). 

\bibitem{Braaten2006U}E. Braaten and H.-W. Hammer, Universality in
few-body systems with large scattering length, Phys. Rep. $\mathbf{428}$,
259 (2006).

\bibitem{Blume2012F}D. Blume, Few-body physics with ultracold atomic
and molecular systems in traps, Rep. Prog. Phys. $\mathbf{75}$, 046401
(2012).

\bibitem{Greene2017U}C. H. Greene, P. Giannakeas, and J. P$\mathrm{\acute{e}}$rez-R$\mathrm{\acute{i}}$os,
Universal few-body physics and cluster formation, Rev. Mod. Phys.
$\mathbf{89}$, 035006 (2017). 

\bibitem{Tan2008E}S. Tan, Energetics of a strongly correlated Fermi
gas, Ann. Phys. $\mathbf{323}$, 2952 (2008).

\bibitem{Tan2008L}S. Tan, Large momentum part of a strongly correlated
Fermi gas, Ann. Phys. $\mathbf{323}$, 2971 (2008).

\bibitem{Tan2008G}S. Tan, Generalized virial theorem and pressure
relation for a strongly correlated Fermi gas, Ann. Phys. $\mathbf{323}$,
2987 (2008).

\bibitem{Zwerger2011T}W. Zwerger, \textit{The BCS-BEC Crossover and
the Unitary Fermi Gas}, Lecture Notes in Physics Vol. 836 (Springer,
Berlin, 2011), see Chap. 6 for a brief review.

\bibitem{Braaten2008E}E. Braaten and L. Platter, Exact relations
for a strongly interacting Fermi gas from the operator product expansion,
Phys. Rev. Lett. $\mathbf{100}$, 205301 (2008).

\bibitem{Zhang2009U}S. Zhang and A. J. Leggett, Universal properties
of the ultracold Fermi gas, Phys. Rev. A $\mathbf{79}$, 023601 (2009).

\bibitem{Stewart2010V}J. T. Stewart, J. P. Gaebler, T. E. Drake,
and D. S. Jin, Verification of universal relations in a strongly interacting
Fermi gas, Phys. Rev. Lett. $\mathbf{104}$, 235301 (2010).

\bibitem{Kuhnle2010U}E. D. Kuhnle, H. Hu, X.-J. Liu, P. Dyke, M.
Mark, P. D. Drummond, P. Hannaford, and C. J. Vale, Universal behavior
of pair correlations in a strongly interacting Fermi gas, Phys. Rev.
Lett. $\mathbf{105}$, 070402 (2010).

\bibitem{Yu2015U}Z. H. Yu, J. H. Thywissen, and S. Z. Zhang, Universal
relations for a Fermi gas close to a $p$-wave interaction resonance,
Phys. Rev. Lett. $\mathbf{115}$, 135304 (2015).

\bibitem{Yoshida2015U}S. M. Yoshida and M. Ueda, Universal high-momentum
asymptote and thermodynamic relations in a spinless Fermi gas with
a resonant $p$-wave interaction, Phys. Rev. Lett. $\mathbf{115}$,
135303 (2015).

\bibitem{He2016C}M. Y. He, S. L. Zhang, H. M. Chan, and Q. Zhou,
Concept of a contact spectrum and its applications in atomic quantum
Hall states, Phys. Rev. Lett. $\mathbf{116}$, 045301 (2016).

\bibitem{Luciuk2016E}C. Luciuk, S. Trotzky, S. Smale, Z. Yu, S. Zhang,
and J. H. Thywissen, Evidence for universal relations describing a
gas with $p$-wave interactions, Nature Phys. $\mathbf{12}$, 599
(2016).

\bibitem{Peng2016L}S.-G. Peng, X. J. Liu, and H. Hu, Large-momentum
distribution of a polarized Fermi gas and $p$-wave contacts, Phys.
Rev. A $\mathbf{94}$, 063651 (2016).

\bibitem{Peng2019U}S.-G. Peng, Universal relations for a spin-polarized
Fermi gas in two dimensions, J. Phys. A: Math. Theor. $\mathbf{52}$,
245302 (2019).

\bibitem{Zhang2017E}P. Zhang, S. Zhang, and Z. Yu, Effective theory
and universal relations for Fermi gases near a $d$-wave-interaction
resonance, Phys. Rev. A $\mathbf{95}$, 043609 (2017).

\bibitem{Landau1999Q}L. D. Landau and E. M. Lifshitz, \textit{Quantum
Mechanics} (Butterworth-Heinemann, Oxford, 1999).

\bibitem{Mott1965T}N. F. Mott and H. S. W. Massey, \textit{Theory
of Atomic Collisions} (3rd ed., Oxford, 1965).

\bibitem{Lin2011S}Y. J. Lin, K. Jiménez-Garcia, and I. B. Spielman,
Spin-orbit-coupled Bose-Einstein condensates, Nature $\mathbf{471}$,
83 (2011). 

\bibitem{Cheuk2012S}L. W. Cheuk, A. T. Sommer, Z. Hadzibabic, T.
Yefsah, W. S. Bakr, and M. W. Zwierlein, Spin-injection spectroscopy
of a spin-orbit-coupled Fermi gas, Phys. Rev. Lett. $\mathbf{109}$,
095302 (2012).

\bibitem{Wang2012S}P. J. Wang, Z. Q. Yu, Z. K. Fu, J. Miao, L. H.
Huang, S. J. Chai, H. Zhai, and J. Zhang, Spin-orbit-coupled degenerate
Fermi gases, Phys. Rev. Lett. $\mathbf{109}$, 095301 (2012).

\bibitem{Wu2016R}Z. Wu, L. Zhang, W. Sun, X. T. Xu, B. Z. Wang, S.
C. Ji, Y. Deng, S. Chen, X. J. Liu, and J. W. Pan, Realization of
two-dimensional spin-orbit coupling for Bose-Einstein condensates,
Science $\mathbf{354}$, 83 (2016).

\bibitem{Huang2016E}L. H. Huang, Z. M. Meng, P. J. Wang, P. Peng,
S. L. Zhang, L. C. Chen, D. H. Li, Q. Zhou, and J. Zhang, Experimental
realization of two-dimensional synthetic spin-orbit coupling in ultracold
Fermi gases, Nat. Phys. $\mathbf{12}$, 540 (2016).

\bibitem{Chen2018S}H.-R. Chen, K.-Y. Lin, P.-K. Chen, N.-C. Chiu,
J.-B. Wang, C.-A. Chen, P.-P. Huang, S.-K. Yip, Y. Kawaguchi, and
Y.-J. Lin, Spin-orbital-angular-momentum-coupled Bose-Einstein condensates,
Phys. Rev. Lett. $\mathbf{121}$, 113204 (2018).

\bibitem{Chen2018R}P.-K. Chen, L.-R. Liu, M.-J. Tsai, N.-C. Chiu,
Y. Kawaguchi, S.-K. Yip, M.-S. Chang, and Y.-J. Lin, Rotating atomic
quantum gases with light-induced azimuthal gauge potentials and the
observation of the Hess-Fairbank effect, Phys. Rev. Lett. $\mathbf{121}$,
250401 (2018). 

\bibitem{Zhang2019G}D. Zhang, T. Gao, P. Zou, L. Kong, R. Li, X.
Shen, X.-L. Chen, S.-G. Peng, M. Zhan, H. Pu, and K. Jiang, Ground-state
phase diagram of a spin-orbital-angular-momentum-coupled Bose-Einstein
condensate, Phys. Rev. Lett. $\mathbf{122}$, 110402 (2019).

\bibitem{Kohler2006P} T. Köhler, K. Góral, and P. S. Julienne, Production
of cold molecules via magnetically tunable Feshbach resonances, Rev.
Mod. Phys. $\mathbf{78}$, 1311 (2006). 

\bibitem{Zhu2006S}S.-L. Zhu, H. Fu, C.-J. Wu, S.-C. Zhang, and L.-M.
Duan, Spin Hall effects for cold atoms in a light-induced gauge potential,
Phys. Rev. Lett. $\mathbf{97}$, 240401 (2006).

\bibitem{Peng2014M}S.-G. Peng, S. Tan, and K. Jiang, Manipulation
of $p$-wave scattering of cold atoms in low dimensions using the
magnetic field vector, Phys. Rev. Lett. $\mathbf{112}$, 250401 (2014).

\bibitem{Gross2017Q}C. Gross and I. Bloch, Quantum simulations with
ultracold atoms in optical lattices, Science $\mathbf{357}$, 995
(2017).

\bibitem{Qi2011T}X. L. Qi and S. C. Zhang, Topological insulators
and superconductors, Rev. Mod. Phys. $\mathbf{83}$, 1057 (2011).

\bibitem{Dalibard2011C}J. Dalibard, F. Gerbier, G. Juzeliunas, and
P. Öhberg, Colloquium: Artificial gauge potentials for neutral atoms,
Rev. Mod. Phys. $\mathbf{83}$, 1523 (2011).

\bibitem{Zhang12019T}D. W. Zhang, Y. Q. Zhu, Y. X. Zhao, H. Yan,
and S. L. Zhu, Topological quantum matter with cold atoms, Adv. Phys.
$\mathbf{67}$, 253 (2019).

\bibitem{Wang2016T}J. K. Wang, W. Yi, and W. Zhang, Two-body physics
in quasi-low-dimensional atomic gases under spin-orbit coupling, Front.
Phys. $\mathbf{11}$, 118102 (2016).

\bibitem{Cui2012M}X. Cui, Mixed-partial-wave scattering with spin-orbit
coupling and validity of pseudopotentials, Phys. Rev. A $\mathbf{85}$,
022705 (2012).

\bibitem{Zhang2012M}P. Zhang, L. Zhang, and Y. Deng, Modified Bethe-Peierls
boundary condition for ultracold atoms with spin-orbit coupling, Phys.
Rev. A $\mathbf{86}$, 053608 (2012).

\bibitem{Peng2018C}S.-G. Peng, C.-X. Zhang, S. Tan, and K. Jiang,
Contact theory for spin-orbit-coupled Fermi gases, Phys. Rev. Lett.
$\mathbf{120}$, 060408 (2018).

\bibitem{Zhang2020U}C.-X. Zhang, S.-G. Peng, and K. Jiang, Universal
relations for spin-orbit-coupled Fermi gases in two and three dimensions,
Phys. Rev. A $\mathbf{101}$, 043616 (2020).

\bibitem{Zhang2018U}P. Zhang and N. Sun, Universal relations for
spin-orbit-coupled Fermi gas near an $s$-wave resonance, Phys. Rev.
A $\mathbf{97}$, 040701(R) (2018).

\bibitem{Jie2018U}J. Jie, R. Qi, and P. Zhang, Universal relations
of an ultracold Fermi gas with arbitrary spin-orbit coupling, Phys.
Rev. A $\mathbf{97}$, 053602 (2018).

\bibitem{Qin2020L}F. Qin, P. Zhang, and P.-L. Zhao, Large-momentum
tail of one-dimensional Fermi gases with spin-orbit coupling, Phys.
Rev. A $\mathbf{101}$, 063619 (2020).

\bibitem{Qin2020U}F. Qin and P. Zhang, Universal relations for hybridized
$s$- and $p$-wave interactions from spin-orbital coupling, arXiv:
2005.04997v1 (2020).

\bibitem{Duan2013U}H. Duan, L. You, and B. Gao, Ultracold collisions
in the presence of synthetic spin-orbit coupling, Phys. Rev. A $\mathbf{87}$,
052708 (2013).

\bibitem{Guan2016S}Q. Guan and D. Blume, Scattering framework for
two particles with isotropic spin-orbit coupling applicable to all
energies, Phys. Rev. A $\mathbf{94}$, 022706 (2016).

\bibitem{Guan2016A}Q. Guan and D. Blume, Analytical coupled-channel
treatment of two-body scattering in the presence of three-dimensional
isotropic spin-orbit coupling, Phys. Rev. A $\mathbf{95}$, 020702(R)
(2017).

\bibitem{Wang2018U}J. Wang, C. R. Hougaard, B. C. Mulkerin, and X.
J. Liu, Ultracold collisions between spin-orbit-coupled dipoles: General
formalism and universality, Phys. Rev. A $\mathbf{97}$, 042709 (2018).

\bibitem{Wu2013S}Y. Wu and Z. Yu, Short-range asymptotic behavior
of the wave functions of interacting spin-$1/2$ fermionic atoms with
spin-orbit coupling: A model study, Phys. Rev. A $\mathbf{87}$, 032703
(2013).

\bibitem{NOTE-PWave}Here, near the $p$-wave resonance, the divergent
term $r^{-1}$ in the $s$-wave channel of \cite{Cui2012M} could
be eliminated by using $\phi\left({\bf r}\right)$. Then our results
agree with that in \cite{Cui2012M}.
\end{thebibliography}
\end{document}